\documentclass[prd,preprint,12pt,nofootinbib]{revtex4}
\usepackage{epsfig}
\usepackage{feynarts}
\begin{document}
\def\D0{\mbox{D\O}}

\renewcommand{\thefootnote}{\fnsymbol{footnote}}

\title{\vspace{3.in}
\LARGE{Decays of the Littlest Higgs $Z_H$ and 
the Onset of Strong Dynamics}}
\author{\Large{John Boersma\footnotemark[1], Andrew Whitbeck\footnotemark[2]}}
\affiliation{Department of Physics and Astronomy,\\
University of Rochester\\
Rochester, NY 14627-0171\\\vspace{2.cm}}

\begin{abstract}

The Little Higgs mechanism, as realized in various models, 
requires a set of new massive gauge bosons, some of which
mix with gauge bosons of the Standard Model.  For a 
range of mixing angles the coupling of gauge bosons to scalars 
can become strong, ultimately
resulting in a breakdown of perturbative 
calculation.  
This phenomenon is studied in the Littlest Higgs 
model, where the approach to strong dynamics is characterized 
by increasing tree-level decay widths of the neutral $Z_H$ boson to lighter gauge 
bosons plus multiple scalars.
These increasing widths suggest a 
distinctive qualitative collider signature 
for the approach to the strong coupling regime 
of large Higgs and other scalar multiplicities. 
In this work we catalog the
kinematically allowed three-body decays of the $Z_H$, 
and calculate the partial width of the  
process $Z_H \to Z_L H H$.  This partial width 
is found to be larger than the comparable two-body decay 
$Z_H \to Z_L H$ for values of the $SU(2)$ mixing angle 
cosine $\theta$ $<0.13$, indicating divergence 
of the Littlest Higgs sigma field expansion at values of 
cosine $\theta$ larger than a simple parametric calculation 
would suggest. 
Additionally, we present analytical expressions 
for all two-body decays of the Littlest Higgs 
$Z_H$ gauge boson, including the effects of 
all final-state masses.  

\end{abstract}

\footnotetext[1]{jboersma@pas.rochester.edu}
\footnotetext[2]{whitbeck@pha.jhu.edu}
\thispagestyle{empty}
\maketitle

\renewcommand{\thefootnote}{\arabic{footnote}}

\section{INTRODUCTION}

A well known formal difficulty of the Standard Model 
of particle physics is
that the Higgs boson mass has quadratically divergent contributions from 
fermion and gauge boson loops which must be cancelled by a `fine-tuned' tree-level
mass term.  This fine-tuning is removed in supersymmetric models by introducing 
boson partners for Standard Model fermions, and fermion partners for Standard Model
bosons.  Cancellations of quadradic divergences then arise from the 
differing signs of bosonic and fermionic loops.  Recently a new class `Little Higgs'
models\cite{reviews}, has emerged in which cancellations occur between paired particles
of the {\it{same}} statistics:  fermions are paired with new, massive fermions, and bosons with 
new massive bosons.  

This Little Higgs mechanism, as realized in various models, 
requires a set of new massive gauge bosons, some of  
which mix with gauge bosons of the Standard Model.  For some 
range of mixing angles the coupling of gauge bosons to scalars 
can become strong, resulting in a breakdown of perturbative 
calculation.  In this work we probe the limits of 
perturbative reliability by examining the decay partial widths of a 
massive gauge boson. 

The prototype Little Higgs model is the `Littlest Higgs'\cite{LH}.  Here  
we study the decays of a new neutral $SU(2)$ gauge boson in the Littlest Higgs model,
the $Z_H$. The presence and properties of this particle are central to the divergence-canceling 
mechanism of the Littlest Higgs model.

A catalog of Feynman rules for the Littlest Higgs model can be found in Reference \cite{Han}, along with 
a number of phenomenological results. A few important corrections can be found in Reference \cite{Buras}. 
Additional rules for scalar couplings can be found in Reference \cite{KR}. Here we will sketch the 
outlines of the model as needed to establish notation and motivate our work.

The enlarged electroweak space of the Littlest Higgs model is the 
symmetric tensor representation of $SU(5)$, which is broken at a scale $f$ by a tensor vacuum 
expectation value to the coset
$SU(5)/SO(5)$.  The result is $24-10=14$ Goldstone bosons, which are parameterized by a $5\times5$ non-linear
sigma field $\Sigma(x)$.  Four of these Goldstone bosons will become the complex Higgs doublet of the Standard
Model, four will become the longitudinal modes of four new massive gauge bosons, and the remaining six will form a 
massive complex scalar triplet.

The $SU(5)$ group contains an $SU(2)_1 \otimes U(1)_1 \otimes SU(2)_2 \otimes U(1)_2$
subgroup, which is gauged, with gauge couplings $g_1, g_1', g_2,$ and $g'_2$, respectively.  The result is eight
gauge bosons, which after diagonalization to mass eigenstates are the gauge bosons of the Standard 
Model, here labeled $A_L$, $Z_L$, $W_L^{\pm}$, and new gauge bosons with masses of order $f$, labeled $A_H$, $Z_H$,
$W_H^{\pm}$.  The mass diagonalization can be described in terms of the mixing angles:

\begin{eqnarray}
c = {\rm{cos}}\:\theta= \frac{g_1}{\sqrt{{g_1}^2+{g_2}^2}} = \sqrt{1-s^2}
\end{eqnarray}
and
\begin{eqnarray}
c' = {\rm{cos}}\:\theta'= \frac{{g'}_1}{\sqrt{{{g'}_1}^2+{{g'}_2}^2}} = \sqrt{1-{s'}^2}.
\end{eqnarray}

With $SU(5)$ represented in this way, quadratic Higgs mass divergences from Standard Model 
gauge boson loops are cancelled by opposite-sign contributions from the new massive gauge bosons, leaving a 
logarithmic divergence.  At some scale above $10$ TeV this logarithmic contribution must also be cancelled 
if fine-tuning is to be avoided, so we do not view the Littlest Higgs model as complete, but instead expect 
at sufficiently high energies additional contributions from an unidentified complete theory, the
so-called ``ultraviolet completion"~\cite{uv}.   
\begin{table}[t]
\begin{tabular}{|c|c|c|}
\hline
Particles&Spin&${\rm{Mass}}$\\ \hline
$\mathbf{\Phi^0, \Phi^P, \Phi^+, \Phi^{++}, \Phi^-, \Phi^{--}}$&$0$&$\frac{\sqrt{2}m_H f}{v (1-(\frac{4v'f}{v^2})^2)^\frac{1}{2}}$\\
\hline
$\mathbf{T,{\bar{T}}}$&$\frac{1}{2}$&$\frac{v}{{m_t}}(\lambda_1 \lambda_2 f)$ \\ \hline
$\mathbf{A_H}$&$1$&${m_Z}{s_w}(\frac{f^2}{5{s'}^2{c'}^2v^2}-1)^\frac{1}{2}$\\ \hline
$\mathbf{Z_H}$&$1$&${m_W}(\frac{f^2}{s^2c^2v^2}-1)^\frac{1}{2}$\\ \hline
$\mathbf{{W_H}^+, {W_H}^-}$&$1$&${m_W}(\frac{f^2}{s^2c^2v^2}-1)^\frac{1}{2}$\\
\hline
\end{tabular}
\caption{\label{lhparticles} New particles in the Littlest Higgs model~\cite{Han}\cite{Buras}
, where $m_W=\frac{gv}{2}$, $m_Z=\frac{gv}{2c_w}$.}
\end{table}

The requirement that Yukawa terms be gauge-invariant restricts the $U(1)_1$ and $U(1)_2$ `hypercharge' 
assignments of the fermions, such that the sum for each fermion equals the Standard Model
hypercharge~\cite{Han}.  The remaining freedom of assignment is limited to one parameter for the quarks, $y_u$,
and one for the leptons, $y_e$.  This freedom is eliminated if we require that all anomalies are cancelled in 
the theory, resulting in the fixed values $y_u = -\frac{2}{5}$ and $y_e = \frac{3}{5}$ \cite{Han}. 
In this work, for simplicity, we always make the anomaly-cancelling choice.

\begin{figure}[t]
\epsffile{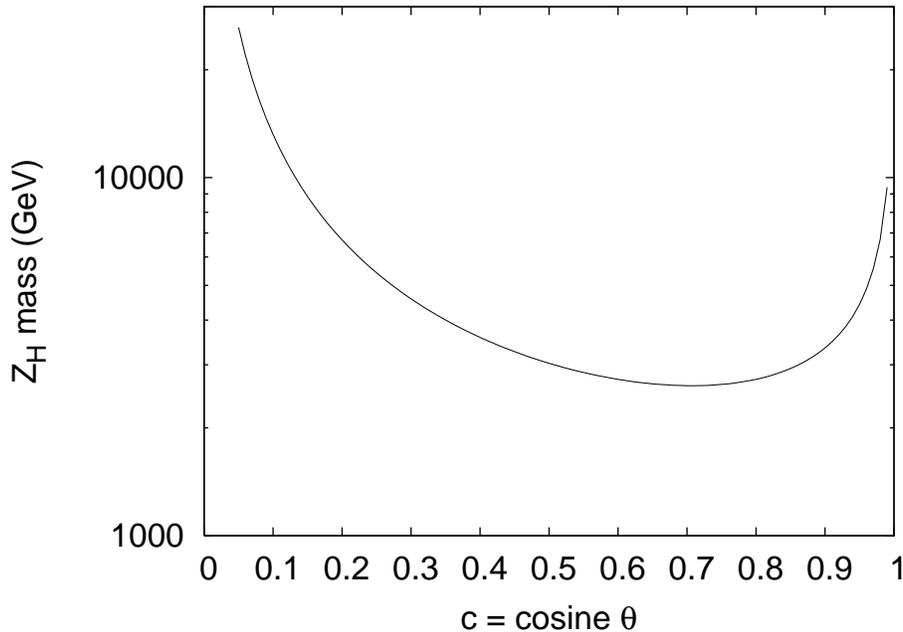}
\caption{\label{zhmass} The mass of the $Z_H$, for $f=4000$ GeV.}
\end{figure}

The new particle content of the Littlest Higgs model is listed in Table \ref{lhparticles}.  All new particles have
masses of order $f$, but not all are of equal importance of an experimental test of this model or one similar to it. 
The presence and couplings of the new $SU(2)$ gauge bosons $Z_H,W^{\pm}_H$ would provide essential tests of the
divergence-canceling mechanism of the model.  The new vectorlike fermion $T$ is necessary to cancel divergences from the 
top quark, but is not structurally related to the rest of the Littlest Higgs model, and so would not have great diagnostic power.
This feature is specific to the Littlest Higgs model; in the Simplest Higgs model \cite{Schmaltz}, for example, the new 
fermion content enters in a less {\it ad hoc} way. 

The $A_H$ is necessary to offset the divergence due to the loop contribution 
from Standard Model hypercharge, but this is numerically negligible, and some variants 
of the Littlest Higgs model have omitted this particle altogether by not gauging the corresponding $U(1)$ group, 
leaving instead an `unconsumed' pseudoscalar\cite{ungauged}. In addition, the properties of the $A_H$ vary greatly
depending on the allocation of hypercharge across the two $U(1)$ groups\cite{myself}.  The couplings of the $Z_H$ are, in contrast, 
highly constrained and central to the structure of the Littlest Higgs model. 

The mass of the $Z_H$ depends on the overall symmetry-breaking scale $f$ and the $SU(2)$ mixing angle, as 
displayed in Figure \ref{zhmass}.  In this
work $f$ is set to $4$ TeV, largely to avoid direct-search exclusion limits already established at the Tevatron\cite{myself}. 
At this scale the mass ranges from a minimum of $2.6$ TeV to in excess of $20$ TeV, depending on the mixing angle.  Since the 
Large Hadron Collider search reach for massive neutral gauge bosons is limited to roughly $5$ TeV\cite{CMS}, the $Z_H$ would not
be experimentally accessible for $c \lesssim 0.27$ at $f=4$ TeV.  The search for the $Z_H$ would in that case have to await an upgraded
LHC or a future, even higher energy collider. 

The free parameters of the Littlest Higgs model and the values used in this study are summarized in Table 
\ref{lhparameters}.    

In Section II of this paper, the two-body decays of the $Z_H$ are described.  Section III
presents three-body decays, compares them to 
two-body decays, and identifies a symptom 
of perturbative breakdown. In Section IV, we summarize our results and draw some conclusions. 

\begin{table}[t]
\begin{tabular}{|c|c|c|}
\hline
\hspace{.4cm} $f$ \hspace{.6cm}&$SU(5)/SO(5)$ symmetry-breaking vev& $4.0$ TeV\\ \hline
$c$&Cosine of the $SU(2)_1 \otimes SU(2)_2$ mixing angle& $0.05$ to $0.99$\\ \hline
$c'$&Cosine of the $U(1)_1 \otimes U(1)_2$ mixing angle& $0.86$\\ \hline
$\lambda_1$&Top sector Yukawa parameter&$1.0$\\ \hline
$y_u$&Quark hypercharge assignment parameter&$-\frac{2}{5}$\\ \hline
$y_e$&Lepton hypercharge assignment parameter&$+\frac{3}{5}$\\ \hline
$v'$&Scalar triplet vev&$1.0$ GeV\\ \hline
\end{tabular}
\caption{\label{lhparameters}  Free parameters of the Littlest Higgs model~\cite{Han}, with the values
used in this study.}
\end{table}

\section{TWO-BODY DECAYS}

The $Z_H$ has a large variety of decays to two particles in the final state.  
In a reliable perturbative expansion we expect these to be the dominant decay 
channels, and so to collectively approximate the total decay width. 
There are decays to all Standard Model fermion-antifermion pairs with
partial widths:
\begin{eqnarray}
\Gamma (Z_H \to f \bar{f} ) = 
\frac{N_c\, m_{Z_H}}{6\pi}\left(1-4\frac{m_f^2}{{m^2_{Z_H}}}\right)^{\frac{1}{2}}
\left(\frac{gc}{4s}\right)^2 \left(1-\frac{m_f^2}{{m^2_{Z_H}}}\right)
\end{eqnarray}
where $N_c$ is the number of fermion colors, $m_{Z_H}$ is the mass of the $Z_H$, 
$m_f$ is the mass of the Standard Model fermion\footnote{
In the numerical results presented below, all Standard Model fermion masses other than the top quark mass
have been omitted since they are utterly negligible compared to the mass of the 
$Z_H$.}, $c$ and $s$ are the cosine
and sine of the mixing angle $\theta$ diagonalizing the $[SU(2)]^2$ gauge bosons,
and $g$ is the Standard Model weak coupling constant.  There are decays to standard 
model $W^+W^-$ pairs with partial width:
\begin{eqnarray}
\Gamma ( Z_H \to W^+W^- ) = 
g^2_{Z_HW_LW_L}\frac{m_{Z_H}}{192\pi}\left(\frac{m_{Z_H}}{m_W}\right)^4
\left(1-4\frac{m^2_W}{m^2_{Z_H}}\right)^\frac{3}{2}
\left(1+20\frac{m^2_W}{m^2_{Z_H}}+12\frac{m^4_W}{m^4_{Z_H}}\right)
\end{eqnarray}
where $m_W$ is the mass of the $W$ boson and the coupling constant is:
\begin{eqnarray}
g_{Z_HW_LW_L}=
\frac{g}{2}\frac{v^2}{f^2}sc(c^2-s^2).
\end{eqnarray}

\begin{figure}[t]
\epsffile{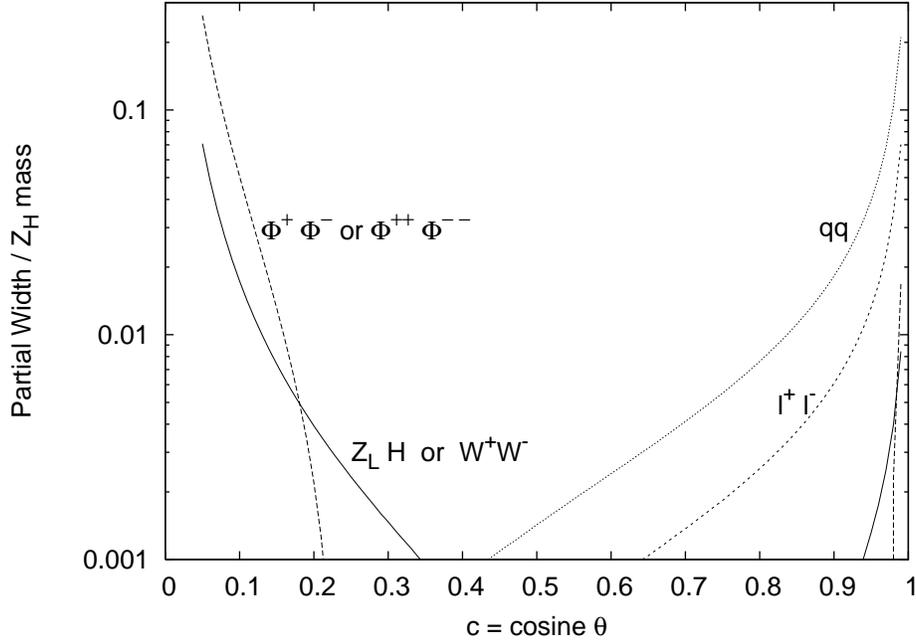}
\caption{\label{zhwtm2f4} The partial width to mass ratio of the $Z_H$
for significant two-body decays, at $f=4$ TeV.
The curve labeled qq is to any Standard Model quark-antiquark pair 
(since even the top mass is negligible at this scale), and the 
curve $l^+ l^-$ is to electrons, and identically to any Standard Model lepton-antilepton pair.}
\end{figure}

Decays to the Standard Model $Z^0$ boson and a Higgs boson occur with 
partial width:
\begin{eqnarray}
\Gamma ( Z_H \to Z^0 H ) = 
\frac{|\vec{P}|V^2}{24\pi m^2_{Z_H}}\left(2+\frac{E^2}{m^2}\right)
\end{eqnarray}
where $|\vec{P}|$,$E$, and $m$ are the fixed momentum, energy, and mass, 
respectively, of the outgoing $Z^0$ in the $Z_H$ rest frame, and the vertex factor constant $V$ is:
\begin{eqnarray}
V = \frac{g^2}{2c_w}\frac{c^2-s^2}{2sc}v,
\end{eqnarray}
with $c_w$ the cosine of the Standard Model weak mixing angle and
$v$ the electroweak scale $v=246$ GeV. 
If the final-state bosons have negligible mass compared to the 
$Z_H$ mass, the partial widths $\Gamma ( Z_H \to W^+W^- ) $ and $\Gamma ( Z_H \to Z^0 H)$ 
reduce to equivalent expressions, as required by the Goldstone boson equivalence theorem, and 
as can be seen in Figure \ref{zhwtm2f4}. 

The $Z_H$ also has decays to final states with one or more particles beyond those in the standard 
model.  Since all the new particle masses in the Littlest Higgs model depend, at least weakly, on 
the free parameters of the theory, a given decay will typically be kinematically possible 
only for some portion of the parameter space.  All the decay branches described here are possible for 
at least some region of parameter space. It is convenient to group these decays according
 to final state spins.  First, decays to a 
vector boson plus a scalar take the same form as to $Z^0\,H$:
\begin{eqnarray}
\Gamma (Z_H \to {\rm{Vector}} + {\rm{Scalar}}) = 
\frac{|\vec{P}|V^2}{24\pi m^2_{Z_H}}\left(2+\frac{E^2}{m^2}\right)
\end{eqnarray}
where $|\vec{P}|$,$E$, and $m$ are the momentum, energy, and mass
of the outgoing vector boson.  The kinematically allowed processes
of this type are $Z_H \to Z^0 \phi^0, A_H \phi^0, W^\pm \phi^\mp$ and $ A_H H$.
The vertex constant for each process is presented in \cite{Han}.

Various decays to two scalars are possible. The general form of this partial
width is:
\begin{eqnarray}
\Gamma (Z_H \to {\rm{Scalar}} + {\rm{Scalar}}) = 
\frac{|\vec{P}|^3}{6\pi m^2_{Z_H}}V^2.
\end{eqnarray}
The possible decays of this type are $Z_H \to H \phi^P, \phi^0 \phi^P, 
\phi^+ \phi^-$, and $\phi^{++} \phi^{--}$. Vertex constants $V$ again are as in \cite{Han}.

Finally, decays to top quark plus the new antifermion $\bar{T}$ (and the conjugate process) 
are possible.  The kinematics of this process are complicated by the differing masses of the 
final state fermions, with result:
\begin{eqnarray}
\Gamma (Z_H \to t \bar{T}) = 
\frac{N_c|\vec{P}|}{6\pi} g^2_v \left(1-\frac{m^2_t}{m^2_{Z_H}}-\frac{m^2_T}{m^2_{Z_H}}
+\left(1-\frac{(m_T-m_t)^2}{m^2_{Z_H}}\right)^2\right)
\end{eqnarray} 
where $g_v$ is the vector coupling (equal and opposite to the axial coupling) for these particles\cite{Han}. 

The partial widths of these various proceses vary greatly with c, as shown in Figure \ref{zhwtm2f4}. At the 
scale $f=4$ TeV, even the effect of the top quark mass is negligible, so decays to top pairs have the same 
width as to other quark-antiquark pairs.  Note that decays to massive $\Phi$ scalars dominate for $c \lesssim 0.2$, 
which has been omitted in some prior studies, perhaps due to concerns about 
perturbativity.  All two-body processes not included in the figure contribute 
a partial width of less that one percent of the mass for all values of $c$.  

\begin{table}
\begin{tabular}{|c|c|}
\hline
$Z_L H H$&All c\\
\hline
$Z_L W^+_L W^-_L$&All c\\
\hline
$A_H H H$&All c\\
\hline
$A_L W^+_L W^-_L$&All c\\
\hline
$A_H W^+_L W^-_L$&All c\\
\hline
$W^+_L H \phi^-$&c $\le 0.49$ or c $\ge 0.88$\\
\hline
$Z_L H \phi^0$&c $\le 0.48$ or c $\ge 0.88$\\
\hline
$A_H H \phi^0$&c $\le 0.38$ or c $\ge 0.93$\\
\hline
$Z_L \phi^0 \phi^0$&c $\le 0.23$ or c $\ge 0.98$\\
\hline
$Z_L \phi^p \phi^p$&c $\le 0.23$ or c $\ge 0.98$\\
\hline
$Z_L \phi^{++} \phi^{--}$&c $\le 0.23$ or c $\ge 0.98$\\
\hline
$A_L \phi^{++} \phi^{--}$&c $\le 0.23$ or c $\ge 0.98$\\
\hline
$W^+_L \phi^0 \phi^-$&c $\le 0.23$ or c $\ge 0.98$\\
\hline
$W^+_L \phi^p \phi^-$&c $\le 0.23$ or c $\ge 0.98$\\
\hline
$W^+_L \phi^+ \phi^{--}$&c $\le 0.23$ or c $\ge 0.98$\\
\hline
$A_H \phi^0 \phi^0$&c $\le 0.20$ or c $\ge 0.98$\\
\hline
$A_H \phi^p \phi^p$&c $\le 0.20$ or c $\ge 0.98$\\
\hline
$A_H \phi^{++} \phi^{--}$&c $\le 0.20$ or c $\ge 0.98$\\
\hline
\end{tabular}
\caption{\label{threes} The kinematically allowed
ranges of the $SU(2)$ mixing angle c, for 
the allowed decays of the Littlest Higgs $Z_H$
to three particle final states, at $f=4$ TeV and 
$c'=0.86$.} 
\end{table}

Although Standard Model final state masses are negligible at $f=4$ TeV, much lower symmetry-breaking scales are 
not excluded by direct searches for some regions of parameter space\cite{myself}.  At $f=1.5$ TeV, the final 
state mass effects are on the order of ten percent. 

\section{THREE-BODY DECAYS}

We surveyed the available Littlest Higgs couplings and particle masses to determine the possible three-body
decays of the $Z_H$.
There are eighteen distinct kinematically-allowed
three-body tree-level decays for the $Z_H$, which are listed in Table \ref{threes}. 
Many of these decay processes are possible only for small or very large values of $c$ (where $Z_H$ becomes very 
massive) due to the presence of one or more TeV-scale particles in the final state.  Note that no decays to the $W_H$ are 
possible because it has the same mass as the $Z_H$ to high order in $\frac{v}{f}$. 

\begin{figure}
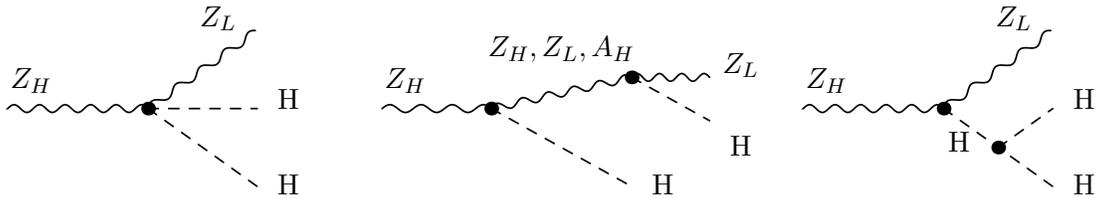

\hspace{-3.5cm}
\unitlength=1.bp
\begin{feynartspicture}(388,140)(3,1)
\FADiagram{}
\FAProp(5.0,10.0)(14.0,10.0)(0.,){/Sine}{0}
\FALabel(6.5,11.0)[b]{$Z_H$}
\FAVert(14.0,10.0){0}
\FAProp(14.0,10.0)(21.0,15.0)(0.,){/Sine}{0}
\FALabel(18.5,15.0)[b]{$Z_L$}
\FAProp(14.0,10.0)(21.0,10.0)(0.,){/ScalarDash}{0}
\FALabel(23.0,10.0)[b]{H}
\FAProp(14.0,10.0)(21.0,5.0)(0.,){/ScalarDash}{0}
\FALabel(23.0,4.5)[b]{H}
\FADiagram{}
\FAProp(7.0,10.0)(14.0,10.0)(0.,){/Sine}{0}
\FALabel(8.5,11.0)[b]{$Z_H$}
\FAVert(14.0,10.0){0}
\FAProp(14.0,10.0)(23.0,12.0)(0.,){/Sine}{0}
\FALabel(18.5,13.0)[b]{$Z_H,Z_L,A_H$}
\FAProp(14.0,10.0)(23.0,5.0)(0.,){/ScalarDash}{0}
\FALabel(25.0,4.5)[b]{H}
\FAVert(23.0,12.0){0}
\FAProp(23.0,12.0)(28.0,9.22)(0.,){/ScalarDash}{0}
\FALabel(30.0,7.0)[b]{H}
\FAProp(23.0,12.0)(28.0,12.0)(0.,){/Sine}{0}
\FALabel(30.0,12.0)[b]{$Z_L$}
\FADiagram{}
\FAProp(12.0,10.0)(21.0,10.0)(0.,){/Sine}{0}
\FALabel(13.5,11.0)[b]{$Z_H$}
\FAVert(21.0,10.0){0}
\FALabel(22.0,7.25)[b]{H}
\FAProp(21.0,10.0)(28.0,15.0)(0.,){/Sine}{0}
\FALabel(25.5,15.0)[b]{$Z_L$}
\FAVert(24.5,7.5){0}
\FAProp(24.5,7.5)(28.0,10.0)(0.,){/ScalarDash}{0}
\FALabel(30.0,10.0)[b]{H}
\FAProp(21.0,10.0)(28.0,5.0)(0.,){/ScalarDash}{0}
\FALabel(30.0,4.5)[b]{H}
\end{feynartspicture}
\caption{\label{Feynmans} The Feynman diagrams contributing 
to the decay process $Z_H \to Z_L H H$.}
\end{figure}

These decays will characteristically each involve a set of eight or so Feynman diagrams, including one with a quadrilinear
vertex factor.  We exclude from our list processes without a quadrilinear coupling and with  
a related kinematically possible two-body decay, since these would constitute a 
double-counting of a two-body decay with an added decay step for one of the decay products. 

The diagrams for the process 
$Z_H \to Z_L H H$ are displayed in Figure \ref{Feynmans}. Note that a diagram 
with a $\phi$ particle propagator is not present because the combined effects of a factor of $\frac{v}{f}$ at 
each vertex places this diagram at higher order in the sigma expansion than we are working. 

\begin{figure}[t]
\epsffile{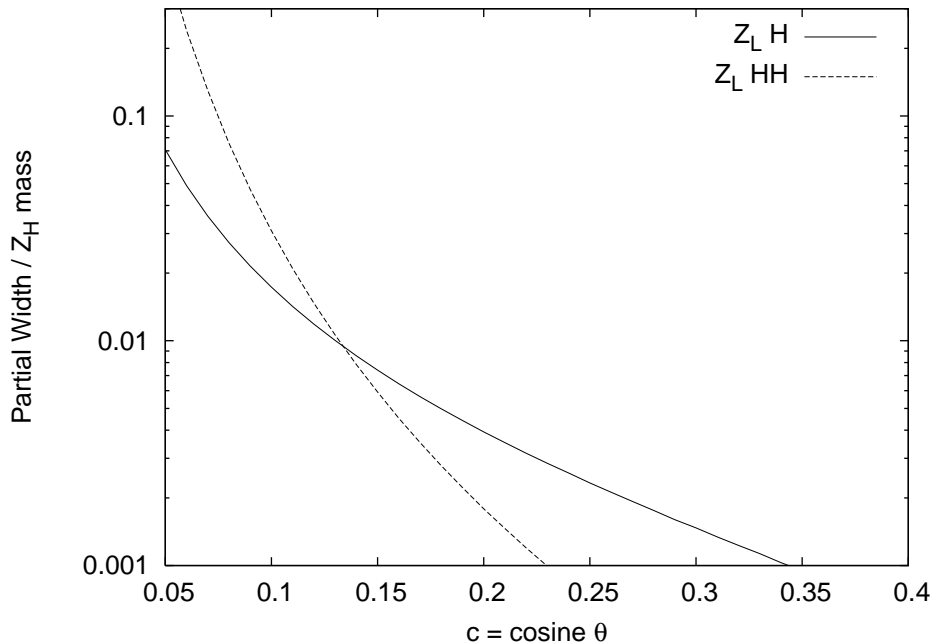}
\caption{\label{threep} The partial width to mass ratio of the $Z_H$
decays to the $Z_L H$ and $Z_L H H$ 
final states, at $f=4$ Tev.  The decay width to $Z_L H H$ exceeds that to 
$Z_L H$ for $c < 0.13$, signalling a breakdown in perturbative reliability.
In the approach to this region, increasing decay widths to multiple scalars 
suggest a distinctive collider signature.}
\end{figure}

Since the calculation of a three-body decay is analytically difficult, especially because we wish to retain 
final state masses, we implemented a Monte Carlo integration to determine the 
partial width for $Z_H \to Z_L H H$. This partial width becomes very large at large  
and especially at small $c$. We would expect this behavior because many vertex factors contain both $c$ and 
$s$ in the denominator. In fact this effect is present in two-body decays, but for three-body decays the 
width divergence is much more severe.  The width-to-mass ratios for the processes $Z_H \to Z_L H$ and $Z_H \to Z_L H H$ are
shown in Figure \ref{threep}.  The three-body partial width exceeds the two-body partial width for $c < 0.13$. 

This is clearly pathological, since three-body decays should be coupling-suppressed 
compared to two-body decays.  
What is happening?  The scalar-gauge boson couplings 
originate in the scalar kinematic Lagrangian term\cite{Han}:

\begin{eqnarray}
\mathcal{L}_\Sigma = \frac{f^2}{8} (\mathcal{D}_\mu \Sigma) (\mathcal{D}^\mu \Sigma)^\dagger
\end{eqnarray}
where $\mathcal{D}_\mu$ is the gauge-covariant derivative in the bifundamental representation of 
$SU(5)$,
\begin{eqnarray}
\label{deriv}
\mathcal{D}_\mu = \partial_\mu \Sigma - i \sum^2_{j=1} 
\left( g_j (W_j\Sigma + \Sigma W^T_j ) + g'_j (B_j \Sigma + \Sigma B^T_j) \right)
\end{eqnarray}
and $\Sigma$ is the
nonlinear representation:
\begin{eqnarray}
\Sigma = e^{2i \Pi / f} \Sigma_0
\end{eqnarray}
and where $\Pi$ is a $5$ x $5$ matrix which
parametrizes the scalars in the theory, including the longitudinal modes
of massive Standard Model gauge bosons as well as the new TeV scale gauge bosons. 
The constant matrix $\Sigma_0$ is the choice of vacuum expectation value for this model\cite{LH}. 
In the gauge-covariant derivative the index $j$ is over the pairs of electroweak 
gauge groups, and the index over the generators of $SU(2)$ are suppressed. 

As $\Sigma$ is expanded in a power series, an infinite set of vertex factors will result from 
the scalar kinematic term, each involving two gauge bosons, some number of scalars and 
a power of $\frac{v}{f s c}$, with the factor $\frac{1}{sc}$ appearing due to mixing 
of the two $SU(2)$ gauge groups.  The lowest 
order terms will be two gauge bosons plus one scalar, then plus two scalars, then three scalars, and 
so on.  The existence of vertex factors connecting more than four external particles signals the fact 
that any nonlinear sigma model is a nonrenormalizable effective theory.  The effects of this nonrenormalizability
would become evident in loop calculations, but here we are considering a different kind of divergence - that of 
the expansion of the $\Sigma$ field itself.  We might expect this divergence to become parametrically significant at 
about $\frac{v}{f s c} \sim 1$, which at $f=4$ TeV occurs at $c=0.06$\cite{Tim}.  Instead, we see the pathological 
effect of the decay to $Z_L H H$ exceeding the decay to $Z_L H$ at as high as $c=0.13$, as we approach the strong coupling 
regime. 

We have not presented the partial widths for other three-body processes, but can note that the decay $Z_H \to 
Z_L W^+_L W^-_L$ should approximate that of $Z_L H H$ due to the Goldstone boson equivalence across these processes. 
With the exception of $Z_H \to A_L W^+_L W^-_L$, all other processes will be kinematically suppressed due to the
presence of one, two, or three TeV-scale particles in the final state.  These processes will in fact be kinematically excluded 
except for small or very large $c$.  By combining these contributions 
we can roughly estimate that the total partial width to three final state particles is less than 
an order of magnitude larger than that to $Z_L H H$ for all but the smallest and largest values of $c$.  

\section{CONCLUSIONS}

We have found that a characteristic three-body decay can become large at 
small and very large values of the $SU(2)$ mixing angle $c$, and 
exceeds a related two-body decay at $c < 0.13$, indicating divergence 
of the sigma field expansion.  The pathology of the Littlest Higgs model
in this region of parameter space is due to a nonlinear sigma scalar field 
representation, and occurs over a larger regions of parameter space than a 
simple parametric calculation would suggest. 

Although perturbative calculations break down in this region, we can note an
interesting qualitative feature:  The Littest Higgs model near the nonperturbative 
region would produce decays with high scalar multiplicities, especially with multiple 
Higgs boson production, and so with large b quark multiplicities.  This feature of large rates 
of scalar production in decays may be a general feature of nonlinear sigma fields near their 
perturbative limits, as the regime of strong dynamics is approached.  In particular, we would expect 
similar results for decays of the Littlest Higgs $A_H$ and $W^\pm_H$, and for the new gauge bosons of 
other models in the Little Higgs class, such as the Simplest Higgs $Z'$. We should note, however, that 
all resuts presented here are at tree-level only, and could be significantly modified by the effects of 
loop corrections.  

We have presented the various expressions for two-body decays of the 
Littlest Higgs $Z_H$, including all final state masses.  The numerical 
width-to-mass ratios for the most significant of these processes, and 
for a characteristic three-body decay process,
confirm that the $Z_H$ is 
a narrow resonance, as needed for dilepton invariant mass searches  
\cite{2006}.

\section*{Acknowledgment}
This material is based upon work supported by the Department of Energy under Award Number
DE-FG02-91ER40685.  Thanks to Dave Rainwater for various helpful discussions.

\end{document}